# AC-feasible Local Flexibility Market with Continuous Trading

Aikaterini A. Forouli*, Georgios K. Papazoglou*,†, Emmanouil A. Bakirtzis*,
Pandelis N. Biskas† and Anastasios G. Bakirtzis†
*Department of Strategy & Business Development, Hellenic Energy Exchange S.A., Athens, Greece
{a.forouli, g.papazoglou, e.bakirtzis}@enexgroup.gr
† Power Systems Lab., School of Electrical & Computer Engineering, Aristotle University of Thessaloniki, Thessaloniki, Greece
{papazoglou, pbiskas, bakiana}@ece.auth.gr

*Abstract*—This paper proposes a novel continuous Local Flexibility Market where active power flexibility located in the distribution system can be traded. The market design engages the Market Operator, the Distribution System Operator and Market Participants with dispatchable assets. The proposed market operates in a single distribution system and considers network constraints via AC network sensitivities, calculated at an initial network operating point. Trading is possible when AC network constraints are respected and when anticipated network violations are alleviated or resolved. The implementation allows for partial bid matching and is computationally light, therefore, suitable for continuous trading applications. The proposed design is thoroughly described and is demonstrated in a test distribution system. It is shown that active power trading in the proposed market design can lead to resolution of line overloads.

## I. INTRODUCTION

The energy transition of the global energy sector is causing deep transformations in the energy systems worldwide. At the root of the energy transition is the effort to reduce carbon emissions, in order to combat the climate crisis [1]. This has manifested through the proliferation of Renewable Energy Sources (RES) and other forms of Distributed Energy Resources (DERs), as well as the electrification of various industries [1]. While the impact of these changes can be observed in most of the aspects of the modern energy systems, it is most apparent in distribution systems.

Traditional power systems assume that electricity is generated by conventional power plants located in the transmission network, which is then transferred to the end customers through the distribution system [2]. Since a large number of DERs is connected to the distribution system, this assumption is being challenged. The distribution system has not been designed to withstand the flows injected by DERs, which can pose operational challenges to the Distribution System Operator (DSO), such as reverse flows, line congestions and voltage violations [3]. The increasing loading of the distribution system, caused by the electrification of human activities further aggravates these challenges. The traditional way of mitigating these issues would be the capital-intensive solution of further network reinforcement [4].

However, technological advancements enable DSOs to use more cost-effective solutions to address these challenges and create new opportunities for the stakeholders. Some of the DERs located in the distribution systems can provide flexibility to the system. Moreover, the expected rollout of smart meters and further adoption of electric vehicles will give rise to flexible consumers [3]. Therefore, Local Flexibility Markets (LFMs) can be created on a distribution system level, through which assets that lie within the distribution system can trade flexibility in order to alleviate or prevent the operational challenges of the DSO.

The concept of LFMs has emerged recently and has received both academic and business interest. Various design approaches for LFMs have been presented, with their main differences being on their timeframes, the type of offered products, their level of network-awareness, and their clearing methodologies.

Implementations of LFMs have been proposed for different timeframes, in accordance with conventional electricity markets. These include long-term LFMs (e.g. [5], [6], [7]), day-ahead LFMs (e.g. [8], [9]), intraday LFMs (e.g. [10]), and real-time LFMs (e.g. [8], [9]). The timeframe of a LFM is related to the type of products it offers. The main differentiation regarding offered products concerns whether flexibility can be traded in the form of capacity or in the form of energy. When traded in the form of capacity, flexibility can be used by DSOs as a tool for network reinforcement deferral [11]. This type of products is usually featured in long-term markets (e.g. [8], [9]), where capacity is reserved with long-term contracts, and can, if needed, be activated closer to real-time [11]. When traded in the form of energy, which is usually the case in day-ahead, intraday and real-time LFMs, flexibility is dispatched at specific time intervals, based on the market time unit resolution (e.g. [10], [12]).

Regarding the level of network-awareness, a LFM can either disregard the network constraints, consider them using

Part of this research is based on the H2020 European Commission Project "FEVER - Flexible Energy Production, Demand and Storage-based Virtual Power Plants for Electricity Markets and Resilient DSO Operation", under grant agreement no. 864537. The responsibility for the content of this paper lies solely with the authors and does not reflect the opinion of the European Commission.

the DC approximation, or consider them using the AC power flow equations, or some method based on relaxed AC optimal power flow. Disregarding the network constraints (e.g. [10], [14]) results in LFMs that are very simple to implement and solve; however, there is no guarantee regarding the feasibility of the resulting schedule. The DC approximation is widely used in the transmission network to reflect the network constraints. Some works (e.g. [12]) have used the DC approximation to reflect the distribution system's constraints. Unlike transmission networks, in the distribution systems the DC approximation fails to accurately reflect the network's constraints, due to the different line characteristics between the two [15], and the fact that it disregards the voltage magnitudes of each node, which at the distribution system can significantly deviate from their nominal values [15], [16]. Incorporating the network's constraints using AC (e.g. [16]) or relaxed-AC (e.g. [17]) techniques is realistic, but it adds a level of complexity on the implementation of the LFM, and can cause problems in its scalability [11].

LFMs can implement an auction clearing mechanism or a continuous trading approach. Most auction-based LFMs feature two-sided auctions, with the objective to maximize social-welfare [18], and the market clearing is performed after the gate closure. Continuous trading markets do not feature gate closures, and clear as soon as a pair of bids matches. According to [12] and [19], continuous markets might be the preferred clearing method for LFMs, at least in their first stages, since they are better suited for the low liquidity conditions they are expected to have, until they reach maturity. Works that explore the capabilities of continuous LFMs can mainly be found in EU and national projects, such as NODES [20], GOPACS [21] and Enera [22].

On the commercial side, interest surrounding LFMs is rising. For example, Piclo is already running an independent marketplace for trading flexibility online called Piclo Flex [6]. In Piclo Flex, DSOs can call for auction-based LFMs based on their needs where participants can offer flexibility in the form of capacity, and receive a premium for its activation. Another example is EPEX SPOT, which announced its intention to launch an LFM [23]. This platform which "has been built specifically for use cases related to flexibility procurement and congestion management" [23], will allow participants to offer flexibility in the form of energy in an auction-based market.

This paper proposes a novel continuous trading LFM where active power flexibility products are traded. The market design engages the Market Operator (MO), the DSO and the Market Participants (MPs). The LFM operates in a single distribution system and is designed to address anticipated grid issues and more specifically line overloading issues. The LFM takes into account the distribution system constraints by using AC sensitivities to determine the acceptance of bids. To the best of our knowledge, this is the first work that explicitly considers AC network feasibility in a continuous trading mechanism. The proposed network feasibility implementation provides the functionality of partial matching of bids which is crucial for the liquidity of the continuous market. The LFM requires minimal information exchange between the DSO and the MO to respect institutional limitations. The implementation is computationally light, therefore, suitable for continuous trading applications. The proposed design is thoroughly described and is demonstrated in a test distribution system.

The paper is organized as follows. Section II presents the LFM design, the concept of addressing line overloading, the calculation of AC sensitivities and their incorporation in the LFM. Section III demonstrates the LFM concept in a small distribution system and Section IV offers some conclusions, and ideas for further research.

## II. MARKET DESIGN

### A. General market framework and actor engagement

The proposed LFM is placed in a market framework that presupposes the operation of a Day-Ahead Market (DAM) by the Market Operator (MO) and the existence of an Independent System Operator on the transmission level. The DAM produces market schedules for the production & generation of the assets of the transmission system, aggregated portfolios and flows for transmission-distribution connection buses, for all Market Time Units (MTUs) of a delivery day D. The assets located in the distribution network participate through aggregators portfolios in the DAM, and the portfolio market schedule is disaggregated (through respective nominations) by the aggregator to individual assets after the DAM results. The proposed LFM is a continuous trading mechanism operated for a specific distribution system, which begins after the conclusion of the DAM auction. The market players can place orders continuously for any MTU of the delivery day D if the trading gate is open. The trading gate opens for all MTUs after DAM closure and closes for each MTU one hour before the actual physical delivery. The proposed framework works for any MTU duration (15, 30, 60 minutes etc.). The trading timeline is depicted in Figure 1. for a delivery day D with hourly MTUs.

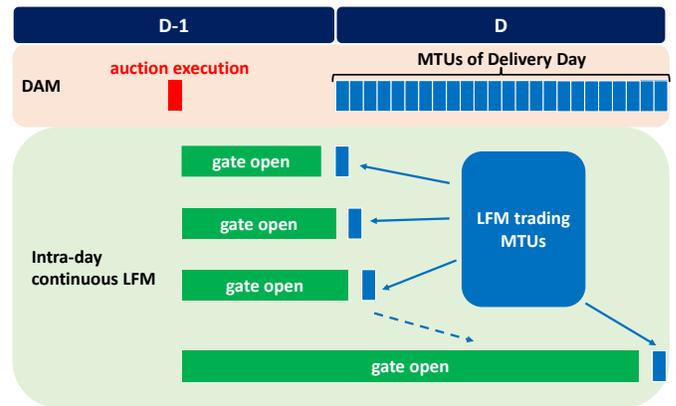

Figure 1. Trading timeline of the intra-day continuous LFM

The proposed LFM enables the continuous trading of active power flexibility in the distribution level. The LFM engages the system operator (DSO), with Market Participants (MPs) and the MO to mitigate anticipated violations of the distribution system physical limitations.

*Market Participants (MPs)*

MPs are participants of the market who represent dispatchable assets, i.e. assets that can alter their generation or consumption based on dispatch instructions, e.g. dispatchable generation, storage, responsive loads etc. MPs submit sell or buy orders for active power based on their available flexibility. Their participation is asset-based, meaning that the submitted order refers to a specific asset located in a specific distribution system node, which has an existing market schedule from the DAM. When a MP's trade is concluded, the market schedule of the affected asset is modified, and so is the expected physical injection or offtake of the asset.

*Distribution System Operator (DSO)*

The DSO after the DAM closure, performs periodically updated forecasts of the non-dispatchable generation and consumption located in the grid for the MTUs of the delivery day. After a forecast is performed, the DSO runs an AC power flow simulation for all forecasted MTUs, identifies anticipated network constraints violations and sends the necessary network data to the MO. The DSO does not submit any orders in the orderbook, and the flexibility requirements are implicitly communicated to the market in the form of remaining available margins of the network operational limits.

*Market Operator (MO)*

The MO is responsible for developing and operating the continuous trading market platform and settlement of transactions. When a new order is submitted, the market platform checks all order combinations that can lead to trades based on a prioritization principle. A concluded trade leads to an update of the market schedule of the relevant MP assets. Orders can be matched upon two conditions:

(a) Financial feasibility: Two orders can be matched only if they abide the financial rule that a buy order can be matched with a sell order only when the buy order price is higher or equal than the sell order price.

(b) Network feasibility: A buy and sell order can be matched only when their trade would lead to a relief of existing network constraint violations, without creating any new ones. The MO receives from the DSO the necessary network data to perform the network feasibility check.

When a trade is concluded, the MO communicates to the MP and the DSO the updated asset market schedule. The DSO runs the AC-power flow and sends the updated network data to the MO.

### B. Market Order characteristics

Orders are submitted by MPs in a standardized format, which must indicate type, direction, location (network bus), MTU for physical delivery, quantity, price and are given a timestamp upon submission. Thus, a typical order is expressed by:

$$O(d, n, mtu, q, p, t)$$

*d*: Order direction (sell/buy)

*n*: Order node location

*mtu*: MTU for which the order is submitted

*q*: Order quantity (MW) during the MTU

*p*: Order price in (€/MW)

*t*: Order submission timestamp (generated by the platform)

MPs submit the available active power flexibility of their assets in the form of buy or sell active power orders. The order price is defined by the bidding strategy employed by the MP. The orders are traded within the trading platform to solve anticipated network constraint violations. In this paper we consider only line overloads as violations.

### C. Grid issues and market mitigation concept

The complex physical laws that describe the operation of power systems dictate that the voltage magnitude of each node, and the apparent power flow of each line, depend on the active and reactive power injections of all the nodes of the network [26]. Trades executed in the proposed market cause changes in the active power injections of the nodes the offers are located, which in turn cause changes in the nodal voltage magnitudes and the line apparent power flows.

The effect of the change of the active power injection of each node on the nodal voltage of each bus and the apparent line flows needs to be quantified, to ensure that the trade contributes in relieving existing violations and does not cause any additional violations of the network constraints. Moreover, due to the continuous nature of the proposed market, this quantification must be done in a fast and efficient way. To this end, the proposed LFM uses AC sensitivities, which can lead to fast and efficient sensitivities calculation, using sparse vector methods [27].

As described in detail in Section E, the sensitivities of the grid's state variables with respect to nodal active power injection are calculated for a specific operating point of the distribution system (i.e. the exact relationship is approximated via a linearization that is performed around the network's initial operating point). Therefore, since the sensitivities are dependent on the network's operating point, they need to be recalculated every time the operating condition of the distribution system changes (e.g. due to more recent forecasts, or to trades being matched). Additionally, the accuracy of the AC sensitivities in calculating the updated network state deteriorates if there is significant deviation between the initial and the final operating points that can be caused if the injections of the assets change significantly (i.e. large quantities of energy are being exchanged). Thus, the use of sensitivities also poses a limitation in the proposed LFM, limiting the maximum quantity of the submitted orders to a predefined level that satisfies the accuracy requirements of the final operating point calculation (which can be set by the DSO). The identification of the maximum bid quantity that the AC sensitivities' accuracy is acceptable is beyond the scope of this paper.

As stated, the DSO periodically performs forecasts and runs the AC power flow to identify anticipated violations of the grid constraints. This paper deals only with line apparent power overload, which essentially refers to a violation in the thermal limit of a line. Line overloads can be addressed in the proposed market design by matching buy and sell active power orders located in different nodes. If the AC sensitivities of the apparent power flow of the overloaded line with respect to the active power injection of buses where the submitted orders are located have the appropriate level and direction, the matching of the orders leads to the alleviation or elimination of the line overload. The AC sensitivities are also used here to ensure that an eligible trade would also not lead to any other grid violations either of the thermal limits of another line or of the voltage magnitude in any bus.

### D. Business process flow

The aforementioned LFM design principles are visualized in the business process flow of Figure 2.

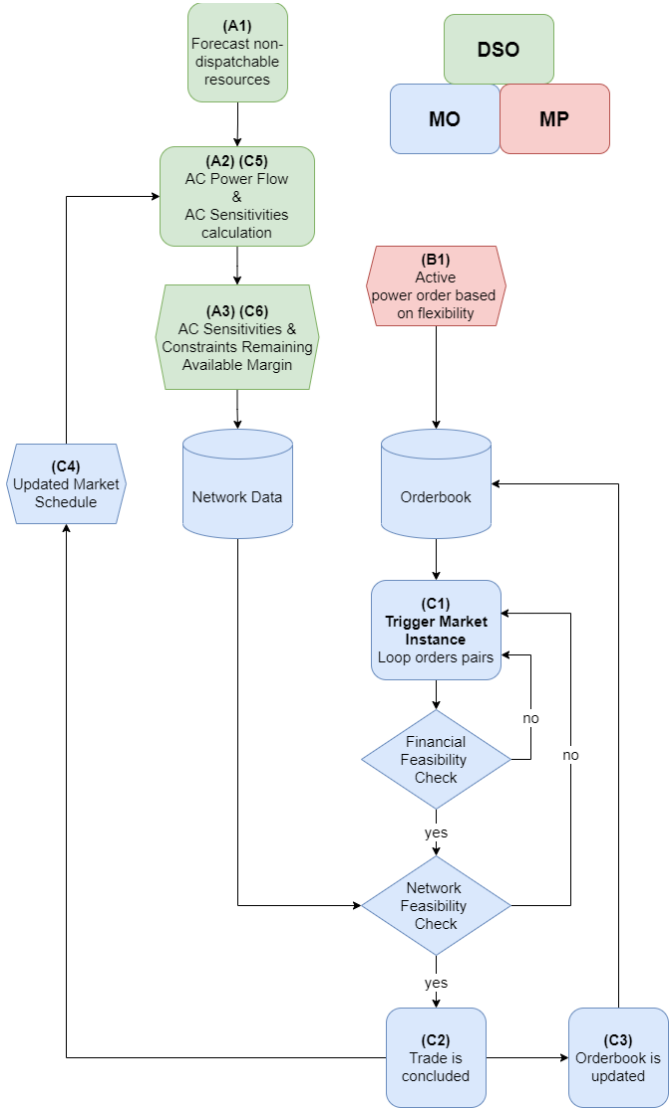

Figure 2. Flow of the market business process

The basic process flow is defined as the market instance and includes the steps $C1-C6$. A market instance is triggered when an order is submitted by a MP in the orderbook ($B1$) or the DSO performs a new forecast ($A1-A3$), and is terminated when all the orders of the orderbook have been checked and no more trades can be concluded or when no violation exists. Within a market instance multiple trades can be concluded. A new market instance for the same MTU can be initiated only when the previous one has been concluded.

The process steps $A1-A3$ are the DSO's actions that can trigger the market instance.

A1. The DSO performs forecasts for the non-dispatchable assets of the distribution system.

A2. The DSO runs an AC power flow and calculates the AC sensitivities of the grid state variables with respect to the nodal active power injections. The computation of the AC sensitivities is presented in Section E.

A3. The DSO sends grid data to the MO. Due to constitutional limitations on information exchange the proposed market design restricts the necessary data that the MO requires to operate the market to:

  a) AC Sensitivities

  b) Remaining available margin of the line operational limits. In case of violation the remaining available margin has a negative value.

The process step $B1$ includes the MP's actions that can trigger the market instance.

B1. While the trading MTU gate is open, MPs can submit active power orders based on their available flexibility.

The process steps $C1-C6$ describe the evolution of a market instance.

C1. A market instance is triggered when an order is submitted by a MP or when the DSO performs a new forecast. The orderbook is sorted based on bid/ask price first, and then on order timestamp. All combinations are checked for matching, starting by checking the orders with the lower sell price and the higher buy price. In case of two orders with the same price and direction, the one with the older timestamp is evaluated first. Naturally, only orders of different direction are eligible for matching. When two orders are checked, the first check is the financial feasibility check, which dictates that a buy order can be matched with a sell order only when the buy order price is higher than or equal to the sell order price. The network feasibility check uses the network data that the MO received by the DSO, to check if the pair of orders that has passed the financial feasibility check can conclude to a trade. The network feasibility checks if the order pair can alleviate or resolve existing violations, without creating new ones. The proposed network feasibility implementation provides the functionality of partial matching of bids which is crucial for the liquidity of the continuous market and is presented in detail in section F.

C2. A trade is concluded when a pair of orders passes both the market and network feasibility checks. The trade is cleared at a single price, which is the price of the order that was first submitted (either the sell or buy order). The relevant MPs are informed about their updated market schedule and are compensated with the trade clearing price. Determination of the trade execution price follows the rules of the European intraday continuous market (incorporated in the XBID project), which imply that two financially feasible orders are matched at the price of the order with the oldest timestamp [28]. Another possible implementation is setting the lowest price between the two orders as the trade execution price. However, alternative pricing mechanisms also exist. For example, in GOPACS market-based flexibility platform [29], non-intuitive trades are allowed (sell price greater than buy price), participants are compensated with pay-as-bid mechanism and the grid operator pays the spread between the buy and sell orders. While it is true that further assessment is required to identify the most appropriate pricing mechanisms for different local flexibility markets, it is evident that the price formation selected in this paper is based on the rules of a well-established operational continuous market and stems upon the principle of providing an advantage to the participant that facilitated first the congestion solution.

- C3. When a trade is concluded, the orderbook is updated. Traded volumes exit the orderbook. If an order has been partially traded, the remaining volume will still be available on the trading platform.
- C4. The MO sends the updated market schedules to the DSO.
- C5. Since a trade is essentially a change in the grid operating point, the DSO runs again an AC power flow for the affected MTU.
- C6. The DSO re-submits the grid data to the MO.

### E. AC Sensitivity Matrices

As stated in Section III C, the DSO runs an AC power flow and calculates the AC sensitivities of line apparent power flow and bus voltage magnitude changes with respect to the bus active power changes. In this section we describe in detail the calculation of the relevant sensitivities.

For the analysis we assume a network with $N$ buses and $M$ lines, where $\Delta P, \Delta Q, \Delta \theta$ and $\Delta U$ are the $N \times 1$ vectors expressing the delta change of bus active and reactive power, voltage angle and magnitude and $\Delta S$ is the $M \times 1$ vector expressing the delta change of line apparent power flow. The proposed LFM uses the following sensitivity relationships:

$$\Delta U = K_{U,P} \, \Delta P \quad (1)$$

$$\Delta S = K_{S,P} \, \Delta P \quad (2)$$

where $K_{U,P}$ is the $N \times N$ sensitivity matrix of the bus voltage magnitude with respect to the bus active power injection, and $K_{S,P}$ is the $M \times N$ sensitivity matrix of the line apparent power flow with respect to the bus active power injection. It is noted that the apparent power flow sensitivities are calculated for both the sending and the receiving ends of each line.

In the proposed LFM $\Delta P, \Delta Q, \Delta \theta, \Delta U, \Delta S$ express the differences between two network states: the initial (known) network state before a trade is concluded, and the final (unknown) network state which is calculated to see if the potential trade can be concluded. It is noted that the initial state is the state where violations are observed. The LFM is a continuous mechanism where only two orders are checked at a time for a potential match. Therefore, in equations (1), (2), only two of the elements of $\Delta P$ are nonzero for each check, which correspond to the nodes where the assets of the orders under consideration lie. For the calculation of the sensitivity matrices the TSO-DSO connection bus is used as a slack bus and will absorb any changes in losses.

*Sensitivities of Nodal Voltage Magnitude with respect to active and reactive nodal power injections*

The steady-state operation of a power system is described by a well-documented system of non-linear power equations, consisting of Kirchhoff's laws and power conservation [26]. A delta change of the network operating point is expressed by:

$$\begin{bmatrix} \Delta P \\ \Delta Q \end{bmatrix} = J \begin{bmatrix} \Delta \theta \\ \Delta U \end{bmatrix} \quad (3)$$

where $J$ is the $2N \times 2N$ Jacobian matrix. Multiplying (3) with the inverse Jacobian matrix, it becomes:

$$\begin{bmatrix} \Delta \theta \\ \Delta U \end{bmatrix} = J^{-1} \begin{bmatrix} \Delta P \\ \Delta Q \end{bmatrix} \quad (4)$$

where the inverse Jacobian matrix $J^{-1}$ is equal to:

$$J^{-1} = \begin{bmatrix} \frac{\partial \theta_1}{\partial P_1} & \frac{\partial \theta_1}{\partial P_2} & \cdots & \frac{\partial \theta_1}{\partial P_N} & \frac{\partial \theta_1}{\partial Q_1} & \frac{\partial \theta_1}{\partial Q_2} & \cdots & \frac{\partial \theta_1}{\partial Q_N} \\ \cdots & \cdots & \cdots & \cdots & \cdots & \cdots & \cdots & \cdots \\ \frac{\partial \theta_N}{\partial P_1} & \frac{\partial \theta_N}{\partial P_2} & \cdots & \frac{\partial \theta_N}{\partial P_N} & \frac{\partial \theta_N}{\partial Q_1} & \frac{\partial \theta_N}{\partial Q_2} & \cdots & \frac{\partial \theta_N}{\partial Q_N} \\ \frac{\partial U_1}{\partial P_1} & \frac{\partial U_1}{\partial P_2} & \cdots & \frac{\partial U_1}{\partial P_N} & \frac{\partial U_1}{\partial Q_1} & \frac{\partial U_1}{\partial Q_2} & \cdots & \frac{\partial U_1}{\partial Q_N} \\ \cdots & \cdots & \cdots & \cdots & \cdots & \cdots & \cdots & \cdots \\ \frac{\partial U_N}{\partial P_1} & \frac{\partial U_N}{\partial P_2} & \cdots & \frac{\partial U_N}{\partial P_N} & \frac{\partial U_N}{\partial Q_1} & \frac{\partial U_N}{\partial Q_2} & \cdots & \frac{\partial U_N}{\partial Q_N} \end{bmatrix} \quad (5)$$

Equation (4) expresses the sensitivity of nodal voltage magnitude and angle with respect to changes of injections of active and reactive power in all the nodes of the distribution system. The sensitivities of the nodal voltage magnitude with respect to changes of injections of active power in all the nodes of the distribution system $K_{U,P}$ is derived from (5) as:

$$K_{U,P} = \begin{bmatrix} \frac{\partial U_1}{\partial P_1} & \frac{\partial U_1}{\partial P_2} & \cdots & \frac{\partial U_1}{\partial P_N} \\ \cdots & \cdots & \cdots & \cdots \\ \frac{\partial U_N}{\partial P_1} & \frac{\partial U_N}{\partial P_2} & \cdots & \frac{\partial U_N}{\partial P_N} \end{bmatrix} \quad (6)$$

*Sensitivities of Line Apparent Power with respect to active nodal power injections*

In order to calculate the sensitivity of apparent power flow changes with regards to changes in nodal active power injections, we first draw from the expressions of the active ($P_{ij}$) and reactive ($Q_{ij}$) power line flows, which for a line starting at node $i$ and ending at node $j$, measured at $i$, are equal to:

$$P_{ij} = U_i^2(g_{ij} + g_{ij}^s) - U_i U_j(g_{ij}\cos\theta_{ij} + b_{ij}\sin\theta_{ij}) \quad (7)$$

$$Q_{ij} = -U_i^2(b_{ij} + b_{ij}^s) - U_i U_j(g_{ij}\sin\theta_{ij} - b_{ij}\cos\theta_{ij}) \quad (8)$$

where $g_{ij}$ and $b_{ij}$ are the line conductance and susceptance, $g_{ij}^s$ and $b_{ij}^s$ are the line shunt conductance and susceptance, $\theta_{ij}$ is the difference between the start bus and end bus voltage angles and $U_i, U_j$ are voltage magnitudes of nodes $i$ and $j$ respectively. The sensitivities of the line apparent power flow $S_{ij}$ with respect to changes in voltages magnitude and angles can then be calculated as:

$$\frac{\partial S_{ij}}{\partial U_n} = \frac{\partial\sqrt{(P_{ij}^2 + Q_{ij}^2)}}{\partial U_n} \quad (9)$$

$$\frac{\partial S_{ij}}{\partial \theta_n} = \frac{\partial\sqrt{(P_{ij}^2 + Q_{ij}^2)}}{\partial \theta_n} \quad (10)$$

where $n$ is either one of the starting node $i$ or ending at node $j$. From (9) and (10), the $M \times 1$ vector of sensitivities for the changes of line apparent power flows, $\Delta S$, with respect to changes in voltages and angles is:

$$\Delta S = J_S \begin{bmatrix} \Delta\boldsymbol{\theta} \\ \Delta\boldsymbol{U} \end{bmatrix} \quad (11)$$

where $J_S$ is the $M \times 2N$ matrix equal to:

$$J_S = \begin{bmatrix} \frac{\partial S_1}{\partial \theta_1} & \frac{\partial S_1}{\partial \theta_2} & \cdots & \frac{\partial S_1}{\partial \theta_N} & \frac{\partial S_1}{\partial U_1} & \frac{\partial S_1}{\partial U_2} & \cdots & \frac{\partial S_1}{\partial U_N} \\ \frac{\partial S_2}{\partial \theta_1} & \frac{\partial S_2}{\partial \theta_2} & \cdots & \frac{\partial S_2}{\partial \theta_N} & \frac{\partial S_2}{\partial U_1} & \frac{\partial S_2}{\partial U_2} & \cdots & \frac{\partial S_2}{\partial U_N} \\ \cdots & \cdots & \cdots & \cdots & \cdots & \cdots & \cdots & \cdots \\ \frac{\partial S_M}{\partial \theta_1} & \frac{\partial S_M}{\partial \theta_2} & \cdots & \frac{\partial S_M}{\partial \theta_N} & \frac{\partial S_M}{\partial U_1} & \frac{\partial S_M}{\partial U_2} & \cdots & \frac{\partial S_M}{\partial U_N} \end{bmatrix} \quad (12)$$

It is noted that based on (6) only four elements of each row have non-zero values.

Combining (11) and (4), we obtain the sensitivity of line apparent power flow with respect to bus active and reactive power injections:

$$\Delta S = J_S J^{-1} \begin{bmatrix} \Delta\boldsymbol{P} \\ \Delta\boldsymbol{Q} \end{bmatrix} \quad (13)$$

where:

$$J_S J^{-1} = \begin{bmatrix} \frac{\partial S_1}{\partial P_1} & \frac{\partial S_1}{\partial P_2} & \cdots & \frac{\partial S_1}{\partial P_N} & \frac{\partial S_1}{\partial Q_1} & \frac{\partial S_1}{\partial Q_2} & \cdots & \frac{\partial S_1}{\partial Q_N} \\ \frac{\partial S_2}{\partial P_1} & \frac{\partial S_2}{\partial P_2} & \cdots & \frac{\partial S_2}{\partial P_N} & \frac{\partial S_2}{\partial Q_1} & \frac{\partial S_2}{\partial Q_2} & \cdots & \frac{\partial S_2}{\partial Q_N} \\ \cdots & \cdots & \cdots & \cdots & \cdots & \cdots & \cdots & \cdots \\ \frac{\partial S_M}{\partial P_1} & \frac{\partial S_M}{\partial P_2} & \cdots & \frac{\partial S_M}{\partial P_N} & \frac{\partial S_M}{\partial Q_1} & \frac{\partial S_M}{\partial Q_2} & \cdots & \frac{\partial S_M}{\partial Q_N} \end{bmatrix} \quad (14)$$

The sensitivity matrices of the line apparent power flow with respect to bus power injection and $K_{S,P}$ is derived from (14) as:

$$K_{S,P} = \begin{bmatrix} \frac{\partial S_1}{\partial P_1} & \frac{\partial S_1}{\partial P_2} & \cdots & \frac{\partial S_1}{\partial P_N} \\ \frac{\partial S_2}{\partial P_1} & \frac{\partial S_2}{\partial P_2} & \cdots & \frac{\partial S_2}{\partial P_N} \\ \cdots & \cdots & \cdots & \cdots \\ \frac{\partial S_M}{\partial P_1} & \frac{\partial S_M}{\partial P_2} & \cdots & \frac{\partial S_M}{\partial P_N} \end{bmatrix} \quad (15)$$

It is again noted that $K_{S,P}$ needs to be calculated for both ends of each line.

### F. Network feasibility check

In this section we describe the network feasibility check process performed by the LFM platform shown in Figure 2. The objective of the LFM is to clear the maximum order quantity that alleviates or eliminates existing violations, without causing violations in other parts of the network. The LFM essentially allows for partial order execution, instead of an all-or-none approach, which increases the liquidity of the market. There are two sets of network operational constraints that are checked: (a) thermal limits of lines and (b) bus voltage magnitude limits.

Assume one buy and one sell active power order, submitted in busses $b$ and $s$, respectively. The effect of a potential trade to the network operational constraints, based on (1)−(2), is expressed as:

$$\Delta S_m = K_{S,P}^{b,m} \cdot \Delta P_b + K_{S,P}^{s,m} \cdot \Delta P_s \quad \forall\, m\, in\, M \quad (16)$$

$$\Delta U_n = K_{U,P}^{b,n} \cdot \Delta P_b + K_{U,P}^{s,n} \cdot \Delta P_s \quad \forall\, n\, in\, N \quad (17)$$

where $\Delta S_m$ is the change in the line apparent power flow of line $m$, $\Delta U_n$ is the change in bus voltage magnitude of bus $n$, $K_{U,P}^{b,n}$ $K_{U,P}^{s,n}$ $K_{S,P}^{b,m}$ $K_{S,P}^{s,m}$ are the sensitivity matrices' elements from (6),(15), and $\Delta P_b$ and $\Delta P_s$ are the order quantities that are executed.

To keep active power balance and assuming buy order has a positive quantity and sell order has a negative quantity, the cleared quantities of both orders should match:

$$\Delta P_b = -\Delta P_s \quad (18)$$

By defining:

$$K_{S,P}^{diff,m} = K_{S,P}^{b,m} - K_{S,P}^{s,m} \quad (19)$$

$$K_{U,P}^{diff,m} = K_{U,P}^{b,m} - K_{U,P}^{s,m} \quad (20)$$

Equations (16) −(17) can be written as

$$\Delta S_m = K_{S,P}^{diff,m} \cdot \Delta P_b \quad \forall\, m\, in\, M \quad (21)$$

$$\Delta U_n = K_{U,P}^{diff,n} \cdot \Delta P_b \quad \forall\, n\, in\, N \quad (22)$$

*Apparent power flow limits of lines*

A potential trade should not cause violations of the line apparent power limits of non-congested lines, thus must respect:

$$S_m^{init} + \Delta S_m \leq S_m^{max} \quad \forall\, m\, \in \{M - M^{viol}\} \quad (23)$$

where $S_m^{max}$ is the thermal limit of line $m$, $S_m^{init}$ is the initial line apparent power flow and $\Delta S_m$ is the change of the line apparent power flow if the eligible trade is concluded, as calculated by (16). Constraint (23) applies for all lines $m$, apart from the line(s) in $M^{viol}$ which is the set of lines with already detected violations. Additionally, a potential trade should be able to reduce or eliminate existing violations in overloaded lines:

$$\Delta S_m < 0 \quad \forall\, m \in M^{viol} \quad (24)$$

Taking into account (21) and the fact that $\Delta P_b$ is always positive, the constraint for any potential trade is:

$$K_{S,P}^{diff,m} < 0 \quad \forall\, m \in M^{viol} \quad (25)$$

An overload is totally eliminated if apart from (24), also (23) is true for the overloaded line. Conclusively, the constraints imposed by the line thermal limits on the maximum accepted order quantity, considering (21),(23),(24), are expressed as:

$$\Delta P_b \cdot K_{S,P}^{diff,m} \leq S_m^{max} - S_m^{init} \quad \forall\, m \in \{M - M^{viol}\} \quad (26)$$

$$\begin{cases} K_{S,P}^{diff,m} < 0 \\ \Delta P_b \geq \dfrac{S_m^{max} - S_m^{init}}{K_{S,P}^{diff,m}} \end{cases} \quad \forall\, m \in M^{viol} \quad (27)$$

In (27) if only the first constraint is true, then we would have a partial alleviation of a line violation.

*Bus voltage magnitude limits*

A potential trade should not cause violations of the bus voltage magnitude limits of buses without violations, thus respecting:

$$U_n^{min} \leq U_n^{init} + \Delta U_n \leq U_n^{max} \quad \forall\, n \in N \quad (28)$$

Combining (22) and (28), the constraints for the quantity range imposed by the bus voltage limits are expressed as:

$$U_n^{min} - U_n^{init} \leq \Delta P_b \cdot K_{U,P}^{diff,n} \leq U_n^{max} - U_n^{init} \quad \forall\, n \in N \quad (29)$$

Ultimately, the quantity range of an eligible active power trade, is constrained by the three equation sets (26),(27),(29).

## III. TEST CASES

In this section we demonstrate how the proposed LFM can mitigate an anticipated line overload by means of a simple numerical example. We use a test distribution system with 5 buses, 5 lines, 3 loads and 3 DERs as illustrated in Figure 3. Bus 0 is considered as the slack bus. The DER generators are modelled with a constant active and reactive power feed-in. The upper and lower bus voltage limits are 1.05pu and 0.95pu respectively, while the line thermal limits are given in TABLE I.

TABLE I. LINE THERMAL LIMITS

| Line index | "From bus" index | "To bus" index | $S_m^{max}$ (MVA) |
|---|---|---|---|
| 0 | 0 | 1 | 12.54 |
| 1 | 1 | 2 | 5.023 |
| 2 | 1 | 3 | 6.755 |
| 3 | 2 | 3 | 5.023 |
| 4 | 4 | 2 | 6.755 |

The initial state of the test distribution system is shown in Figure 3. It can be observed that line 1 is overloaded.

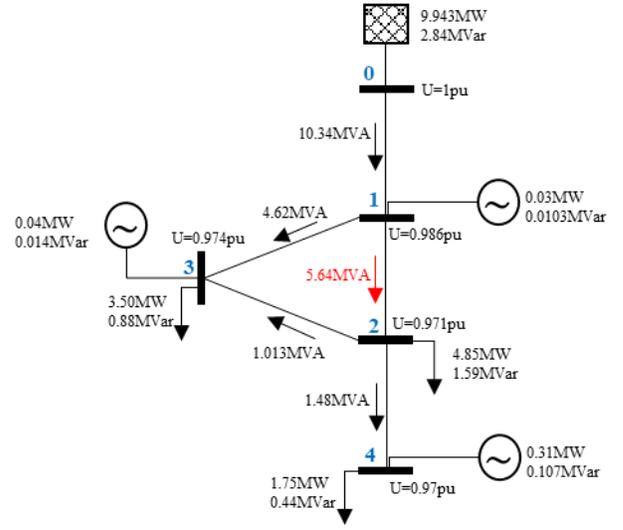

Figure 3. Initial state of a network with line overload

Assuming the LFM gate is open for orders submission, and without loss of generality, we consider two orders of opposite direction that have passed the financial feasibility check, a buy order of 3 MW submitted by the dispatchable load in bus 3 and a sell order of 2 MW submitted by the dispatchable load in bus 2. We want to find the feasible range of the quantity that can be cleared between the two orders, so as the overload of line 1-2 could be alleviated or eliminated, also ensuring that no further violations would be created. Since the cleared quantity should be the same for the two orders, we will examine if the quantity of 2 MW can solve the congestion and if so, whether it can be fully or partially cleared. The sensitivity factors of the apparent power flow at both ends of all lines, to the active power changes in buses 3 and 2, are depicted in TABLE II.

TABLE II. LINE APPARENT POWER FLOW SENSITIVITY FACTORS

| Line index | | Sensitivity factors $K_{S,P}$ | | $K_{S,P}^{diff,m}$ |
|---|---|---|---|---|
| | | Bus 3 | Bus 2 | |
| From bus | 0 | 1.014 | 1.020 | -0.006 |
| | 1 | 0.329 | 0.660 | -0.331 |
| | 2 | 0.657 | 0.332 | 0.325 |
| | 3 | -0.310 | 0.311 | -0.621 |
| | 4 | $-3.55 \cdot 10^{-15}$ | $-4.44 \cdot 10^{-15}$ | $8.88 \cdot 10^{-16}$ |
| To bus | 0 | 0.985 | 0.991 | -0.006 |
| | 1 | 0.320 | 0.639 | -0.319 |
| | 2 | 0.642 | 0.324 | 0.318 |
| | 3 | -0.315 | 0.316 | -0.631 |
| | 4 | $8.70 \cdot 10^{-6}$ | $1.18 \cdot 10^{-5}$ | $-3 \cdot 10^{-6}$ |

The sensitivity factors of the voltage magnitude of each bus to active power changes in buses 3 and 2 shown in TABLE III.

TABLE III. VOLTAGE MAGNITUDE SENSITIVITY FACTORS

| Bus index | Sensitivity factors $K_{S,U}$ | |
|---|---|---|
| | Bus 3 | Bus 2 |
| 0 | 0 | 0 |
| 1 | 0.001214925 | 0.001224315 |
| 2 | 0.001914775 | 0.002598323 |
| 3 | 0.002569388 | 0.001926367 |
| 4 | 0.001917429 | 0.002601924 |

Applying the values of TABLE I. TABLE II. TABLE III. and Figure 3. to equations, (26), (27), (29) for the line and for the voltage magnitude limits, the following limitations apply:

TABLE IV. ACCEPTABLE QUANTITY RANGE OF TRADE

| Network element index | Equation | Lower Limit | Upper Limit |
|---|---|---|---|
| Line apparent power flow limits at "from bus" | 0 (26) | -399.57 | |
| | 1 (27) | **1.87** | |
| | 2 (26) | | **6.57** |
| | 3 (26) | -6.45 | |
| | 4 (26) | | $5.9463 \cdot 10^{15}$ |
| Line apparent power flow limits at "from bus" | 0 (26) | -430.95 | |
| | 1 (27) | 1.68 | |
| | 2 (26) | | 6.88 |
| | 3 (26) | -6.37 | |
| | 4 (26) | -1,703,065.15 | |
| Bus voltage limits index | 0 (29) | | |
| | 1 (29) | -6779.12 | 3870.51 |
| | 2 (29) | -114.99 | 31.30 |
| | 3 (29) | -37.73 | 117.78 |
| | 4 (29) | 116.8 | 29.30 |

The limits shown in TABLE IV. impose a maximum order quantity of 6.57 MW and a minimum order quantity of 1.87 MW to fully resolve the congestion and ensure that no further line or voltage constraint will be violated. If the submitted quantity is less than 1.87 MW, and since the first constraint of (27) is true, the orders will be feasible and thus fully accepted. This is because even if the cleared quantity does not fully solve the overload, it reduces the overload and thus leads to a better condition for the grid. In case where the submitted quantity is greater than 6.57 MW , the orders will be cleared up to the quantity that is within the feasible range, meaning up to the maximum limit of 6.57 MW. This is because a quantity greater than 6.57 MW will lead to a new overload in line 2, as it can be easily understood from the respective limitation of TABLE IV. Owing to the above analysis, the quantity of 2 MW can be fully cleared, and it is expected to fully resolve the congestion. To test this, we rerun the power flow, considering a load increase of 2 MW at node 3 and an equal load decrease at node 2. The results are shown in Figure 4. where indeed the overload of line 1 has been resolved.

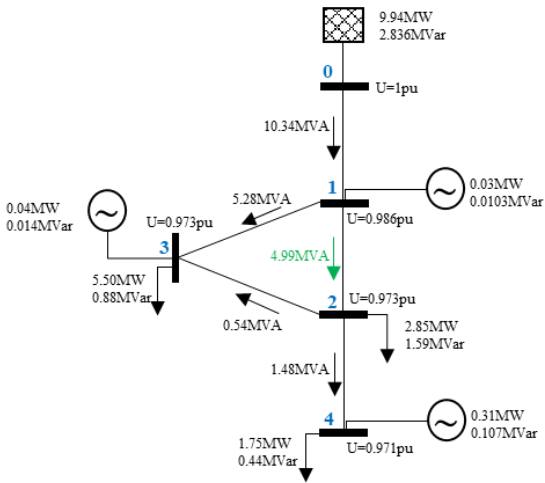

Figure 4. Network state after the trade execution

## IV. CONCLUSIONS

The paper presented a continuous trading LFM where active power flexibility products are traded. The LFM operates in a single distribution system and considers network constraints via AC network sensitivities to address network issues. The proposed idea has been demonstrated in a test case where active power trading can lead to resolution of line overloading issues. Ideas for further research include the consideration of other distribution system violations e.g. voltage violations, deviations in the TSO-DSO schedules as well as a robust mechanism to address multiple simultaneous violations of different nature. We also intend to further analyze the limitations of using AC sensitivities to traded quantities as well as the effect of trading to grid losses. Last, the effects of an unbalanced three-phase distribution system should also be investigated.